\documentclass{Interspeech}

\interspeechcameraready

\title{An approach to measuring the performance of Automatic Speech Recognition(ASR) models in the context of Large Language Model(LLM) powered applications}

\author[affiliation={1}]{Sujith}{Pulikodan}
\author[affiliation={2}]{Sahapthan}{K}
\author[affiliation={3}]{Prasanta Kumar }{Ghosh}
\author[affiliation={1}]{Visruth }{Sanka}
\author[affiliation={1}]{Nihar}{Desai}

\affiliation{AI \& Robotics Technology Park(ARTPARK)}{I-Hub @ IISc}{India}
\affiliation{Department of Mathematics}{ Indian Institute of
Science}{India}
\affiliation{Department of Electrical Engineering}{ Indian Institute of
Science}{India}

\email{sujith@artpark.in, sahapthank@iisc.ac.in, prasantg@iisc.ac.in, visruth@artpark.in, nihar@artpark.in}

\usepackage{xcolor}

\usepackage{comment}
\usepackage{float}

\begin{document}

\maketitle
\begin{abstract}   
Automatic Speech Recognition (ASR) plays a crucial role in human-machine interaction and serves as an interface for a wide range of applications. Traditionally, ASR performance has been evaluated using Word Error Rate (WER), a metric that quantifies the number of insertions, deletions, and substitutions in the generated transcriptions. However, with the increasing adoption of large and powerful Large Language Models (LLMs) as the core processing component in various applications, the significance of different types of ASR errors in downstream tasks warrants further exploration. In this work, we analyze the capabilities of  LLMs to correct errors introduced by ASRs and propose a new measure to evaluate ASR performance for LLM-powered applications. 
\end{abstract}

\begin{table*}[!htbp]
\centering
\begin{tabular}{|l|l|l|l|l|}
\hline
\textbf{Model Name}  & \textbf{Dataset} & \textbf{WER} & \textbf{GPT-4o-mini } & \textbf{ Gemini 1.5 Flash } \\
\hline  
\hline
openai/whisper-tiny                & fleurs        & 14.81\%        & \textbf{12.25\%}                                    & 12.59\%      \\  
openai/whisper-base                & fleurs        & 11.93\%        & \textbf{08.54\%}                                    & 09.30\%      \\  
openai/whisper-small               & fleurs        & 07.62\%        & \textbf{07.30\%}                                    & 07.60\%      \\  
openai/whisper-medium                & fleurs        & \textbf{05.86\%}        & 05.87\%                                    & 06.46\%      \\  
openai/whisper-large                & fleurs        & \textbf{05.64\%}       & 06.01\%                                    & 06.58\%      \\  
facebook-wav2vec2-base-960h                & fleurs        & 20.46\%        & 11.56\%                                    & \textbf{11.42\%}   \\    
facebook-wav2vec2-large-960h                & fleurs        & 17.84\%        & \textbf{10.23\% }                                   & 10.50\%                                      \\
\hline
openai/whisper-tiny                & librispeech        & 08.72\%     & \textbf{08.40\%}                                    & 08.68\%     \\ 
openai/whisper-base                & librispeech        & \textbf{06.37\% }       & 06.64                                   & 06.99\%      \\ 
openai/whisper-small                & librispeech        & \textbf{04.83\%}        & 05.34                                    &06.06\%      \\ 
openai/whisper-medium                & librispeech        & \textbf{04.60\%}        & 05.36                                   & 05.73\%      \\ 
openai/whisper-large                & librispeech        & \textbf{04.41\%}        & 05.20                                   & 26.94\%      \\ 
facebook/wav2vec2-base-960h                & librispeech        & \textbf{03.38\% }       & 04.60                                    & 04.82\%      \\ 
facebook/wav2vec2-large-960h                & librispeech        & \textbf{02.76\% }       & 04.33\%                                    &04.57\%      \\ 
\hline
openai/whisper-tiny                & voxpopuli        & 44.05\%      & 21.54\%                                    & \textbf{21.08\% }     \\ 
openai/whisper-base                & voxpopuli        &36.91\%        & \textbf{19.15\% }                                 & 23.19\%      \\ 
openai/whisper-small                & voxpopuli        & 38.48\%        & 21.83\%                                    &\textbf{21.02\%}      \\ 
openai/whisper-medium                & voxpopuli        & 36.63\%        & \textbf{22.45\% }                                  & 23.43\%      \\ 
openai/whisper-large                & voxpopuli        & 18.96\%        & \textbf{18.46\%}                                & 19.26\%      \\ 
facebook/wav2vec2-base-960h                & voxpopuli        & 20.33\%        & \textbf{12.38\% }                                  & 13.63 \%      \\ 
facebook/wav2vec2-large-960h                & voxpopuli        & 17.80\%        & \textbf{11.32\%}                                    &12.39\%      \\ 

\hline
\end{tabular}
\caption{WER Before and After LLM Correction for Different Models and Datasets}
\end{table*}

\section{Introduction}

Automated Speech Recognition (ASR) has become a cornerstone of human-computer interaction, enabling more natural and seamless communication with machines. Significant advancements in the field have led to increasingly accurate models, some of which even surpass human-level performance \cite{xiong2018microsoft}. However, these achievements are predominantly confined to languages with abundant training data. This disparity highlights considerable biases in the availability and performance of ASR models across different languages, underscoring the need for more inclusive development efforts \cite{zhao2022improving}.

Word Error Rate (WER) is a commonly used metric for benchmarking the performance of ASR models. It evaluates accuracy by accounting for insertions, deletions, and substitutions in the transcriptions. Typically, the selection of an ASR model for a specific application depends on a trade-off between model complexity and performance. When choosing a model, requirements guide this balance to ensure optimal performance while managing computational and resource constraints.

WER has some  limitations, particularly in its inability to capture semantic context and its inadequacy in assessing downstream task performance \cite{kim2021semantic}. To address these shortcomings, various alternative metrics have been proposed. Semantic Word Error Rate (SWER) \cite{spiccia2016semantic}, Word Information Lost (WIL) and Word Information Preserved (WIP) \cite{morris2004and} offer refined assessments by incorporating contextual and informative word preservation. Additionally, embedding-based metrics such as BERTScore \cite{zhang2019bertscore} and Semantic Distance (SemDist)\cite{kim2021semantic} leverage deep contextual representations to evaluate ASR outputs more effectively.

Recent advancements in artificial intelligence and hardware technologies have greatly accelerated the development of Large Language Models (LLMs). Trained on vast amounts of data, LLMs now serve as the backbone of many voice-enabled applications, enabling advanced language comprehension. Their strong language understanding capabilities make them highly effective across a wide range of applications \cite{naveed2023comprehensive}. Combining LLMs with ASR systems enables the development of more sophisticated and effective voice-based applications.

In this work, we investigate ASR performance evaluation in the context of LLM-based applications. We first analyze how effectively ASR errors can be mitigated using LLMs’ language understanding capabilities through simple prompts. To quantify this impact, we measure the extent of improvement achieved. Additionally, we introduce a novel metric designed to assess ASR performance within LLM-driven applications. This metric utilizes a more advanced LLM as a judge to evaluate ASR output quality, ensuring a more contextual and semantically aware assessment.

\section{Related work}
Several approaches have been explored to improve ASR performance by incorporating LLMs. S. Li et al.  fine-tuned a multilingual LLM covering more than 100 languages to correct 1-best hypothesis errors from various speech foundation models \cite{li2024investigating}. Z. Ma et al.  introduced SLAM-ASR, which features a frozen speech encoder and a frozen LLM, with a trainable linear projector to align speech and text modalities \cite{ma2024embarrassingly}. R. Sachdev et al.  proposed an approach where an N-best list of hypotheses and a prompt are input into an LLM for error correction \cite{sachdev2024evolutionary}. Similarly, R. Ma  explored both fine-tuning and zero-shot error correction methods using ASR N-best lists, proposing multiple decoding strategies \cite{ma2024asr}. M. Si  introduced the Lexical Error Guard (LEG), leveraging LLMs' pre-trained knowledge and instructional learning to create an adaptable ASR error correction system \cite{si2024lexical}. In another study, R. Ma et al.  evaluated LLMs, specifically ChatGPT, for ASR error correction, demonstrating improvements across transducer and attention-based ASR models on multiple test sets \cite{ma2023can}. Additionally, Pranay Dighe et al.  explored prompt engineering to explain N-best lists to LLMs and fine-tuned Low-Rank Adapters (LoRA) for downstream tasks, effectively integrating ASR outputs with LLMs to enhance performance \cite{dighe2024leveraging}. These studies collectively highlight the potential of LLMs in refining ASR outputs through fine-tuning, prompt engineering, and error correction strategies.

\section{Our Approach}
In this work, we investigate the correction capabilities of LLMs using a one-shot approach, considering that in many use cases, the full n-best output from the ASR system may not be accessible. Our goal is to explore the potential of LLMs by systematically evaluating multiple models across a diverse set of English-language datasets. The evaluation aims to determine whether the performance differences between ASR models remain consistent after applying LLM-based corrections compared to their original outputs.

As part of the work, we propose an alternative metric WER for evaluating ASR performance in LLM-powered applications. This new measure prioritizes understanding context and identifying important words rather than achieving perfect accuracy for every word in the transcription.
\subsection{Error correction by LLMs}

In this part, we aim to harness the capabilities of LLMs to directly correct errors in ASR output. The transcription generated by the ASR system is passed through the LLM, guided by a carefully designed prompt to assist in correcting the errors. A simple prompt is used to refine the ASR output effectively. The prompt is designed using an iterative and test-driven approach, where each iteration is validated on a set of examples through manual inspection. After multiple iterations, the final prompt used for evaluation is provided below. 
\\
\\
\textbf{Prompt used for error correction:} \textit{Carefully review the provided English Automatic Speech Recognition(ASR) output and correct any errors caused by insertion, deletion, or substitution of words.
Ensure that your corrections are strictly made at the word level without altering the structure or phrasing of the original sentence. Your response should consist solely of the corrected text, with no additional explanations, comments, or formatting.
}
\\

The corrected transcription is then evaluated by comparing it with the reference output, and the WER is calculated to quantify the improvement achieved through this approach.

\subsection{LLMs based QA Approach}

In this approach, we evaluate ASR system quality by examining the context-understanding capabilities of LLMs in relation to ASR output. Specifically, we analyze how context comprehension differs when LLMs are presented with the reference text versus the ASR-generated transcription.
\begin{figure}[h]
    \centering
\includegraphics[width=0.45\textwidth]{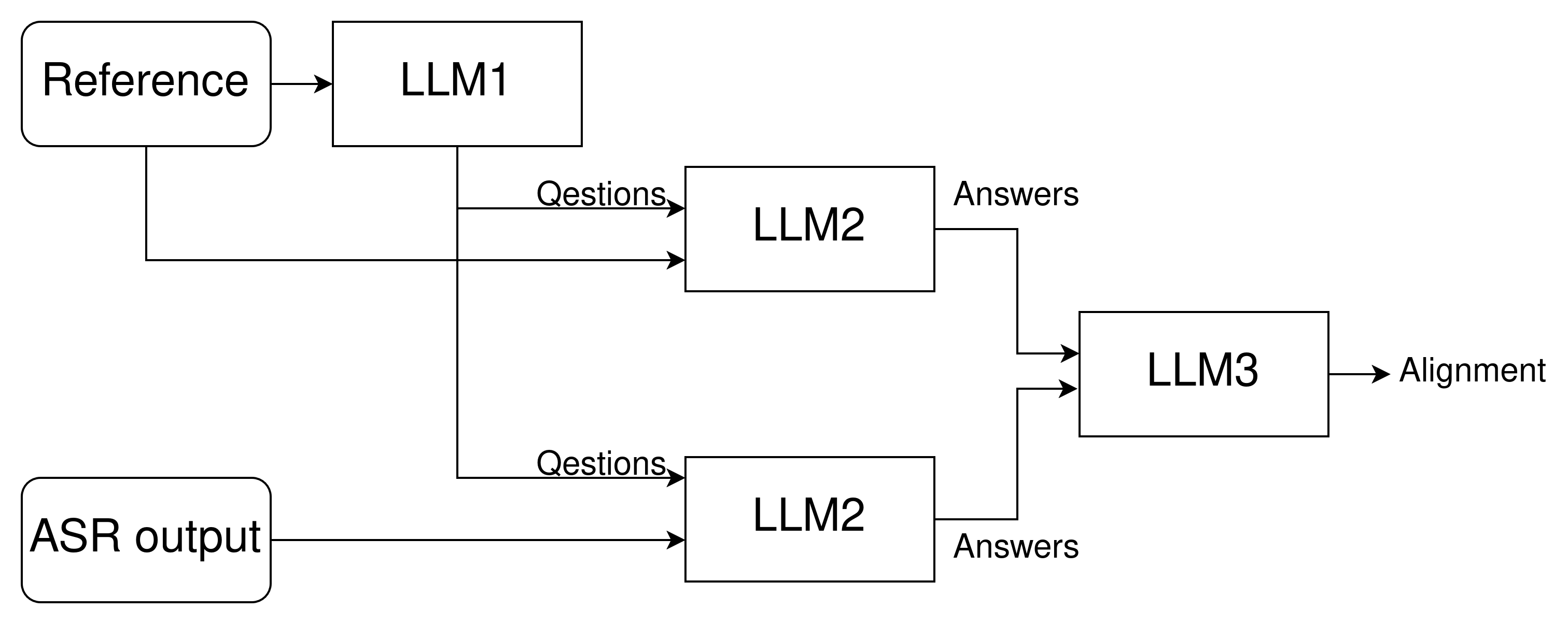}
    \caption{Proposed Evaluation Framework.}
    \label{fig:example}
\end{figure}

The experiment follows a question-and-answer framework. The process starts by generating as many meaningful questions as possible using an LLM(LLM1). These questions are derived from the reference (ground truth), and the LLM is prompted to generate numerous questions based on the provided context. The prompt is carefully designed to ensure that the questions are relevant to the context and maximize the number of meaningful questions generated. The prompt is developed through an iterative, test-driven process, with each version carefully tested on a set of examples and manually reviewed for accuracy and effectiveness.
\\
\\
\textbf{Prompt used for generating questions:} \textit{Create as many questions as possible from the given English text while strictly following these guidelines:\\
Each question must have a answer  derived from the context. Questions should be contextually meaningful and not just based on the words in the sentence. Only include meaningful and relevant questions in English script. Ensure all key aspects of the context are covered through the questions.\\
Sample output format \\
"Questions": [\\
    \{ \\
        "1": "What were the people primarily focused on regarding the individual mentioned in the text?"\\
    \} , \\
    \{
        "2": "What aspect of the individual's character was not understood by those around him?"\\
    \} ]\\
Context:  The Pyramids' Salad and Light Show is one of the most interesting things in the area of kids.
}\\
\\
\begin{table*}[!htbp]
\centering
\begin{tabular}{|l|l|l|l|l|l|l|l|l|l|}
\hline
\textbf{Model Name}  & \textbf{Dataset} & \textbf{WER}& \textbf{LLM1} & \textbf{LLM2} & \textbf{LLM3}     & \textbf{AER} \\
\hline
\hline
openai/whisper-tiny                     & Fleurs       & 14.81\%  &Gemini 1.5 Flash &GPT-4o-mini &GPT-4o                                                                                                          & 24.78\%                                    \\
openai/whisper-base                     & Fleurs       & 11.93\%  &Gemini 1.5 Flash &GPT-4o-mini &GPT-4o                                                                                                           & 20.15\%                                    \\

openai/whisper-small                     & Fleurs       & 07.62\%  &Gemini 1.5 Flash &GPT-4o-mini &GPT-4o                                                                                                            & 15.86\%                                    \\
openai/whisper-medium                     & Fleurs       & 05.86\%  &Gemini 1.5 Flash &GPT-4o-mini &GPT-4o                                                                                                             & 15.08\%                                    \\
openai/whisper-large                     & Fleurs       & 05.64\%  &Gemini 1.5 Flash &GPT-4o-mini &GPT-4o                                                                                                            & 14.77\%                                    \\
facebook/wav2vec2-base-960h           & Fleurs        & 20.46\%    &Gemini 1.5 Flash &GPT-4o-mini &GPT-4o                                                                                                            & 23.93\%                                    \\
facebook/wav2vec2-large-960h           & Fleurs        & 17.84\%    &Gemini 1.5 Flash &GPT-4o-mini &GPT-4o                                                                                                             & 23.01\%                                    \\
\hline
openai/whisper-tiny                     & librispeech       &08.72\%  &Gemini 1.5 Flash &GPT-4o-mini &GPT-4o                                                                                                           & 30.09\%                                    \\
openai/whisper-base                     & librispeech       & 06.37\%  &Gemini 1.5 Flash &GPT-4o-mini &GPT-4o                                                                                                          & 28.97\%                                    \\

openai/whisper-small                     & librispeech       & 04.83\%  &Gemini 1.5 Flash &GPT-4o-mini &GPT-4o                                                                                                             & 26.00\%                                    \\
openai/whisper-medium                     & librispeech       & 04.60\%  &Gemini 1.5 Flash &GPT-4o-mini &GPT-4o                                                                                                            & 26.37\%                                    \\
openai/whisper-large                     & librispeech       & 04.41\%  &Gemini 1.5 Flash &GPT-4o-mini &GPT-4o                                                                                                             & 25.40\%                                    \\
facebook/wav2vec2-base-960h           & librispeech        & 03.38\%    &Gemini 1.5 Flash &GPT-4o-mini &GPT-4o                                                                                                             & 26.96\%                                    \\
facebook/wav2vec2-large-960h           & librispeech        & 02.76\%    &Gemini 1.5 Flash &GPT-4o-mini &GPT-4o                                                                                                           & 26.43\%                                    \\
\hline
openai/whisper-tiny                     & voxpopuli       & 44.05\%  &Gemini 1.5 Flash &GPT-4o-mini &GPT-4o                                                                                                             & 47.23\%                                    \\
openai/whisper-base                     & voxpopuli      &36.91\%  &Gemini 1.5 Flash &GPT-4o-mini &GPT-4o                                                                                                           & 45.05\%                                    \\

openai/whisper-small                     & voxpopuli       & 38.48\%  &Gemini 1.5 Flash &GPT-4o-mini &GPT-4o                                                                                                            & 45.67\%                                    \\
openai/whisper-medium                     & voxpopuli      & 36.63\%  &Gemini 1.5 Flash &GPT-4o-mini &GPT-4o                                                                                                            & 47.39\%                                    \\
openai/whisper-large                     & voxpopuli       & 18.96\%   &Gemini 1.5 Flash &GPT-4o-mini &GPT-4o                                                                                                            & 43.45\%                                    \\
facebook/wav2vec2-base-960h           & voxpopuli        & 20.33\%    &Gemini 1.5 Flash &GPT-4o-mini &GPT-4o                                                                                                              & 43.87\%                                    \\
facebook/wav2vec2-large-960h           & voxpopuli         & 17.80\%    &Gemini 1.5 Flash &GPT-4o-mini &GPT-4o                                                                                                             & 43.66\%                                    \\
\hline
\end{tabular}
\caption{Evaluation of Models with Proposed Approach}
\end{table*}

In the second step, the generated questions are used to obtain answers from the target LLM (LLM2) under evaluation. Answers are generated independently by providing two distinct contexts: the reference text and the ASR output. Each set of answers is generated separately to minimize any potential interference between the two contexts. This process results in two sets of answers for the same set of questions: one based on the reference text and the other based on the ASR output.\\\\
\textbf{Prompt used for generating answers:} \textit{Generate answers for the given questions based on the provided context.
Provide the answers in the form of a dictionary, where the question is the key and the answer is the value, enclosed in brackets. 
Ensure the answers are derived directly from the provided context.
Please provide the the dictionary as the response and don not provide any other content or explanations 
Questions:\\
What is the speaker making?\\
How long does it take to make the magic powder?\\
is the speakers current status on making the powder?\\
For whom is the speaker making the magic powder?\\
What does Margolotte intend to use the magic powder for?\\
What word describes the speakers feeling about the progress of making the powder?\\
What is the implied nature of the magic powder?\\
What is the relationship between the speaker and Margolotte?\\
Context:\\
It takes me several years to make this magic powder but at this moment i am pleased to say it is nearly done you see I am making it for my good wife margolott who wants to use some of it for a purpose of her own\\
}

To evaluate performance, we propose a novel metric termed Answer Error Rate (AER). AER is defined as the ratio of the number of questions with contextually differing answers to the total number of questions. Unlike traditional exact-match metrics, AER focuses on semantic and contextual discrepancies between answers.

To determine whether two answers are contextually different or equivalent, we employ a more powerful LLM (LLM3), guided by a carefully crafted prompt to ensure nuanced comparisons.

In the final step, the two sets of answers—those generated by LLM2 using the ASR output and those generated from the reference text—are compared. Ideally, the answers from the ASR transcription should align closely with those from the clean reference. Discrepancies between them are quantified using AER, which ranges from 0 to 1, with higher values indicating greater divergence in contextual meaning.
\\\\
\textbf{Example Prompt for comparing Answers:} \textit{    You are provided with a dictionary where each key represents a question, and its value is a dictionary containing two answers in the format {'Answer1': value1, 'Answer2': value2}. 
Your task is to determine whether both answers for each question are identical.If the two answers matches( if meaning is essentially the same), mark the result as True; otherwise, mark it as False. 
Return a list of flags where each flag corresponds to the result for a question.
Please provide the output in list format and do not provide anything else\\
Input:
\{
What is the primary activity described in the text?:\{ Answer1: collecting huge amounts of data on citizens, Answer2: collecting huge amounts of data on citizens\}, 
What type of data is being collected?: \{ Answer1: detailed profiles, Answer2: detailed profiles \} , 
Who is collecting the data?: \{ Answer1: not specified, Answer2: not specified\}
\}
}

\section{Datasets and Models used}

For our experiment, we utilized three datasets: Fleurs \cite{conneau2023fleurs}, VoxPopuli \cite{wang2021voxpopuli}, and Librispeech \cite{panayotov2015librispeech}. Specifically, we used 645 utterances from Fleurs, 245 utterances from VoxPopuli, and 250 utterances from Librispeech, all in English. The Fleurs dataset and LibriSpeech both consist of read speech, where speakers deliver pre-written text in a clear and controlled manner. VoxPopuli features spontaneous speech recorded during political proceedings, capturing natural speaking styles with real-world variability. It also includes non-native English speech.

For our evaluations, we utilized two families of ASR models: OpenAI Whisper\cite{radford2022whisper} and Facebook's Wav2Vec2\cite{baevski2020wav2vec20frameworkselfsupervised}, both of which are open source. Specifically, we evaluated five variants of Whisper: Tiny (39M parameters), Base (74M parameters), Small (244M parameters), Medium (769M parameters), and Large (1550M parameters). For Wav2Vec2, we evaluated two variants: Base (95M parameters) and Large (317M parameters).

We employed GPT-4o, GPT-4o-mini \cite{achiam2023gpt}, and Gemini 1.5 Flash \cite{team2024gemini} for our analysis. These are powerful commercial models available in the market, ranking among the leading LLMs in various performance benchmarks \cite{chiang2024chatbotarenaopenplatform}\cite{berkeley-function-calling-leaderboard}.

\section{Results}
The performance of ASRs, along with the error correction approach of LLMs, is presented in Table 1. As expected, the performance of ASR improves with an increase in the number of parameters in models that use the same backend architecture. For datasets consisting of read speech, the performance is relatively higher, as anticipated. However, for the VoxPopuli dataset, which features spontaneous speech, the ASR performance is comparatively lower.

When analyzing the WER after applying error correction by the LLMs, we observe a significant improvement when the ASR's initial WER is on the higher side. This demonstrates the ability of LLMs to leverage their language understanding capabilities to correct errors introduced by the ASR. However, at lower initial WERs, the WER occasionally increases slightly after applying corrections. This may be due to the LLMs attempting to make changes to the text even when minor grammatical or other inconsistencies exist in the reference itself.

In our analysis using two commonly used LLMs, GPT-4.0 Mini and Gemini 1.5 Flash, we observe that the performance gains are comparable. This similarity could be attributed to the reported capabilities of these LLMs, which are closely aligned in benchmarks such as reasoning tasks.

\begin{figure}[h]
    \centering
\includegraphics[width=0.45\textwidth]{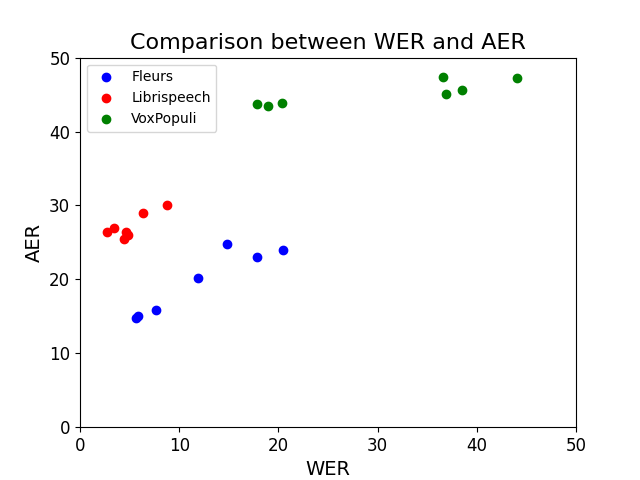}
    \caption{Comparison between WER and AER}
    \label{fig:example}
\end{figure}
The performance comparison of the proposed AER is presented in Table 2. The first observation is that the AER is significantly higher than the WER. This is because AER evaluates the context-awareness of LLMs, while WER impacts context-awareness differently. Errors in important words have a greater impact on AER, as they are critical to maintaining context, whereas WER treats all words equally, regardless of their contextual significance.


We can also observe that the performance gaps in the ASR do not reflect the same gaps in AER. This may be due to the nature of the errors made by the ASR. The ASR tends to make errors in less frequent words, and this issue persists across different ASRs. While LLMs are capable of correcting these words, they often fail to do so for less frequent words. As illustrated in the accompanying plot, a lower Word Error Rate (WER) does not always correspond to a lower AER. We argue that AER provides a better reflection of the ASR system's ability to capture meaningful context in downstream tasks. This has important implications for model selection: relying solely on WER may be insufficient, as models with lower WER do not necessarily translate to improved end-to-end performance. In some cases, larger ASR models with better WER show only marginal gains compared to models with higher WER
\section{Conclusion and Future works}
We can observe that LLMs have the ability to correct ASR output, and we found that a low WER does not necessarily translate to better performance in LLM-based systems. Extending this analysis to multiple LLMs would be useful to better understand the overall trends across different models. 
\section{Acknowledgments}
We sincerely thank the leadership teams of the Indian Institute of Science (IISc) and ARTPARK for their invaluable support, which has enabled us to work on this problem.
\bibliographystyle{IEEEtran}
\bibliography{mybib}

\end{document}